\documentclass[aps,preprint,prd,showpacs,nofootinbib]{revtex4}
\usepackage{amsmath}
\usepackage{graphicx}
\usepackage{dcolumn}
\usepackage{bm}
\usepackage{amssymb}
\usepackage{latexsym}

\def\be{\begin{equation}}
\def\ee{\end{equation}}
\def\ba{\begin{eqnarray}}
\def\ea{\end{eqnarray}}

\begin{document}

\title{Phantom Inflation and Primordial Perturbation Spectrum }

\author{Yun-Song Piao$^{a,b}$}
\email{yspiao@itp.ac.cn}
\author{Yuan-Zhong Zhang$^{c,d}$}
\affiliation{${}^a$Institute of High Energy Physics, Chinese
Academy of Science, P.O. Box 918-4, Beijing 100039, P. R. China}
\affiliation{${}^b$Interdisciplinary Center of Theoretical
Studies, Chinese Academy of Sciences, P.O. Box 2735, Beijing
100080, China} \affiliation{${}^c$CCAST (World Lab.), P.O. Box
8730, Beijing 100080} \affiliation{${}^d$Institute of Theoretical
Physics, Chinese Academy of Sciences, P.O. Box 2735, Beijing
100080, China }

\date{Mar. 23$^\mathrm{th}$, 2004}

\begin{abstract}
In this paper we study the inflation model driven by the phantom
field. We propose a possible exit from phantom inflation to our
observational cosmology by introducing an additional normal scalar
field, similar to hybrid inflation model. Then we discuss the
primordial perturbation spectra from various phantom inflation
models and give an interesting compare with those of normal scalar
field inflation models.

\end{abstract}

 \maketitle

Due to the central role of primordial perturbations on the
formation of cosmological structure, it is important to probe
their possible nature and origin. The basic idea of inflation is
simple and elegant \cite{GLS, LAS}. Though how embedding the
standard scenario of inflation in realistic high energy physics
theory has received much attention (for a recent review see Ref.
\cite{LIN}), exploring new types or classes inflation models is
still an interesting issue.

Phantom matter is being regarded as one of interesting
possibilities describing dark energy \cite{C}, in which the
parameter of state equation $w< -1$ and the weak energy condition
is violated, which is motivated and may be favored by the recent
supernova data \cite{ASSS, CP, A}. The simplest implementing of
phantom matter is a normal scalar field with reverse sign in its
dynamical term. Though the quantum theory of such a field may be
problematic \cite{CHT, CJM}, this dose not mean that the phantom
matter is unacceptable. The actions with phantom-like behavior may
be arise in supergravity \cite{N}, scalar tensor gravity
\cite{BEPS}, higher derivative gravity theories \cite{P}, brane
world \cite{SS}, k-field \cite{ADM} and others \cite{CL, S}.
Stringy phantom energy has also been discussed \cite{FM}. Recently
there are many relevant studies of phantom cosmology \cite{PR, G,
PZ, GD, ST}.

In general phantom matter can drive a superinflation phase, in
which when the scale factor expands, the Hubble parameter also
increases gradually, thus the initial perturbation in the horizon
will exit the horizon, and reenter the horizon after the
transition to an expanding phase which may be regarded as our
observational cosmology in which the Hubble parameter decreases
gradually. Depending on the matching conditions of perturbations,
the nearly scale-invariant spectrum can be generated only during
slow expanding and inflation \cite{PZ}. In this paper we shall
study some aspects of phantom inflation, such as various models,
possible exits, primordial perturbation spectrum, especially focus
on its different features from those of normal scalar field
inflation.

{\it \bf Basic Model -} We start with such an effective action of
simple phantom field as follows \be {\cal L}= {1\over
2}(\partial_{\mu}\varphi)^2 - V(\varphi) \label{mat} \ee where the
metric signature $(-~+~+~+)$ is used. If taking the field
$\varphi$ spatially homogeneous but time-dependent, the energy
density $\rho$ and pressure $p$ can be written as \be \rho
=-{1\over 2}{\dot \varphi}^2 + V(\varphi) ~~~~~ p=-{1\over 2}{\dot
\varphi}^2-V(\varphi) \label{p} \ee From (\ref{p}), for $\rho
>0$ the state parameter $\omega \equiv {p\over \rho}< -1$ can be
seen.

We minimally couple the action (\ref{mat}) to the gravitational
action. The Friedmann universe, described by the scale factor
$a(t)$, satisfies the equation \be h^2 = {8\pi G\over 3}
\left(-{1\over 2 }{\dot \varphi}^2 +V(\varphi)\right) \label{da}
\ee and the dynamical equation of phantom field is \be {\ddot
\varphi} +3h{\dot \varphi} - V^\prime(\varphi)=0 \label{drho} \ee
where $G$ is the Newton gravitational constant and $h ={{\dot
a}\over a}$ is the Hubble parameter. Different from normal scalar
field, in general the phantom field will be driven up along its
potential. As has been analytically and numerically shown in Ref.
\cite{ST} that for asymptotically power-law potential, in late
time analogous to the slow-roll regime for the normal scalar
field, the phantom field will enter into ``slow-climb" regime. For
generic phantom potentials, we assume that the phantom field is
initially in the bottom of potential, then is driven by its
potential and climbs up alone the potential. After defining the
slow-climb parameters \be \epsilon_{\rm pha} \equiv -{{\dot
h}\over h^2 } ~~~~~~ \delta_{\rm pha}\equiv -{{\ddot \varphi}\over
{\dot \varphi}h}\label{ed}\ee when the conditions $|\epsilon_{\rm
pha}|\ll 1$ and $|\delta_{\rm pha}| \ll 1$ are satisfied,
(\ref{da}) and (\ref{drho}) can be rewritten as \be h^2 = {8\pi
G\over 3} V(\varphi) \label{h2}\ee and \be 3h{\dot \varphi} -
V^\prime(\varphi)=0 \label{dp2}\ee Thus we obtain $a\sim e^{ht}$
approximately, {\it i.e.} the universe enters into the
inflationary phase driven by the phantom field, which will
continue up to ``Big Rip" after some finite/infinite time
dependent various phantom potentials \cite{ST}.

{\it \bf Possible Exits -} For possible ``bounce" to the
observational cosmology, we may expect that at some time the
energy of phantom field could be transited into the usual
radiation by reheating \cite{KLS, FKL1, FKL2}, similar to that of
normal scalar fields inflation. But for slow-climbing phantom
field, the reheating is hardly possible by the decay of phantom
field. Even if it is possible, the energy of produced radiation
decays like $a^{-4}$ and the potential energy will dominate
quickly again, thus inflation will restart and does not end.
Through the singular ``Big Rip" by some other
mechanism\footnote{In Ref. \cite{GD}, the ``Big Rip" can be
circumvented by using growing wormholes and ringholes, which
connects the phantom phase and late-time expanding phase.} from
high energy/dimension theory may be regarded as a reasonable
``bounce" to an observational cosmology \cite{BM}. But the
primordial perturbation evolving through the singular ``Big Rip"
will reduce to many uncertainties for late-time observations.

However, a reasonable exit then reheating may be implemented by an
additional normal scalar field $\sigma$, which is very similar to
the case of hybrid inflation \cite{L, CLL, GLW}. We write the
effective potential \footnote{The coupling between a phantom field
and a normal scalar field should be introduced in a cautious
manner. Recently, this issue has received increased attention
\cite{CHT, CJM}. However, instead of the open problems of
(quantum) field theory of phantom, we pay much attention to the
next things, in particular some possible phenomena of early
universe with phantom field. } with phantom field $\varphi$ and
normal scalar field $\sigma$ as \be V(\varphi,\sigma)
=\left(V(\varphi) +\lambda \sigma^2\varphi^2 \right)
e^{-\alpha\sigma^2/\sigma_*^2} \label{v}\ee where $\lambda,
\alpha$ are dimensionless parameter. The aim of multiplying $
e^{-\alpha\sigma^2/\sigma_*^2}$ is to make $V(\varphi,\sigma)=0$
for $\sigma\rightarrow\infty$ and arbitrary $\varphi$, which will
reduce the climb-up motion of $\varphi$ during and after
reheating. Further, to make $V(\varphi)$ not evolve to infinity,
which will make universe enter ``Big Rig", $V(\varphi)$ should
have finite maxima. To briefly illustrate how to exit, we take
$V(\phi)=\Lambda e^{-\beta \varphi^2/\varphi_*^2} $, in which
$\beta$ is dimensionless parameter. Its figure is plotted in Fig.
1. We can see that the minima of this potential are in
$\varphi\rightarrow \infty$, $\sigma =0$ and $\sigma \rightarrow
\infty$ for arbitrary $\varphi$. Initially, we assume that the
value of the phantom field $\varphi$ is very large and
$\sigma\simeq 0$, thus the mass of $\sigma$ field \be m_{\sigma}^2
\simeq 2\lambda \varphi^2-{2\alpha \Lambda\over \sigma^2_*}
e^{-\beta\varphi^2/\varphi^2_*}\simeq 2\lambda \varphi^2
\label{m}\ee is also very large, which will compel the departure
for $\sigma =0$ very small, In this case the evolution of universe
can be driven by only the motion of phantom field $\varphi$ and
follows the equations (\ref{da}) and (\ref{drho}), and the phantom
field $\varphi$ will be driven up along its potential, and enter
into ``slow-climb" regime in late-time and satisfy the equations
(\ref{h2}) and (\ref{dp2}). Its value decreases gradually. After
$m_{\sigma}^2 =0$ is satisfied, {\it i.e.} \be \varphi^2_c \simeq
{\alpha \Lambda\over \lambda \sigma^2_* }\exp(-\beta
\varphi^2_c/\varphi^2_*) \ee the $\sigma$ field will become
unstable, and roll down along the direction of $\sigma >0$ or
$<0$.

\begin{figure}[t]
\begin{center}
\includegraphics[width=8cm]{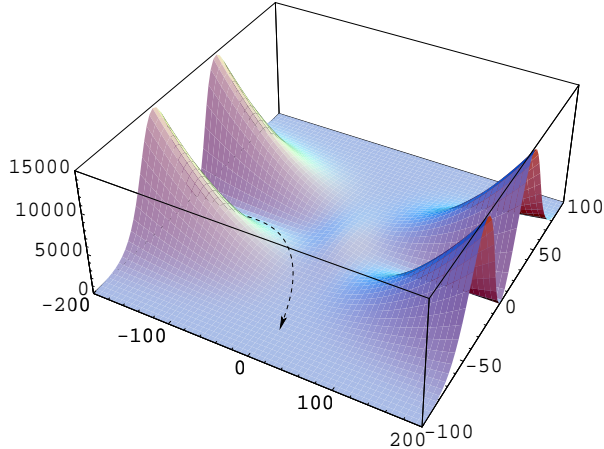}
\caption{ The figure of effective potential (\ref{v}):
$-200<\varphi <200$ and $-100 <\sigma <100$. We take
$\alpha=\beta=\lambda=10^{-3}$, $\Lambda=10^3$ and
$\varphi_*=\sigma_* =1$. In this case, $\varphi_c\simeq \pm 24$.
The dashing line is one of the motive trajectory of
$(\varphi,\sigma)$. } \label{fig1}
\end{center}
\end{figure}

With the increasing of $|\sigma|$ value, the effective potential
gradually decreases. Since $\sigma$ is the normal scalar field,
its coupling to usual matter is well-defined. In our case, the
$\sigma$ field does not oscillate \footnote{Though we select a
run-away potential of the $\sigma$ field, a potential leading to
the oscillation of $\sigma$ after the exit can be also found. Some
further details can be seen in Ref. \cite{PFZ}. }, thus the
standard theory of reheating based on the decay of the oscillating
field \cite{KLS, GL} does not apply. The gravitational particle
production \cite{F} may be a reheating approach but is not
efficient enough. Furthermore, the instant preheating mechanism
\cite{FKL1, FKL2} may be better selection and be more efficient in
such non-oscillatory models.

In this scenario, initially the $(\varphi,\sigma)$ is in the
bottom of the effective potential, then slow climbs along its
valley. During this period, the fluctuations are stretched to
outside of horizon and form primordial perturbations on the
formation of cosmological structure. A generic potential of simple
phantom field relative to this phase can be characterized by two
independent mass scales and be written as \be V(\varphi) =\Lambda
f({\varphi\over \varphi_*})\label{v2}\ee where the height
$\Lambda$ corresponds the vacuum energy density during phantom
inflation and is fixed by normalization, and The width $\varphi_*$
corresponds the change of the field value and is a free parameter
of models. We will discuss the metric perturbations of various
phantom inflation model from such general potential of phantom in
the following.

{\it \bf Primordial Perturbations -} In longitudinal gauge and in
absence of anisotropic stresses, the scalar metric perturbation
can be written as \be ds^2 = a^2(\eta) (-(1+2\Phi)d\eta^2
+(1-2\Phi) \delta_{ij}dx^i dx^j )\ee where $\eta$ is conformal
time $ d\eta ={dt\over a}$ and $\Phi$ is the Bardeen potential.
Defining the variable \cite{M, MFB} \be v
={a\varphi^\prime\over {\cal H}} \zeta \equiv z\zeta\ee
where \be \zeta= {1\over \epsilon}({\cal H} \Phi^\prime+\Phi)+\Phi
\ee is the curvature perturbation on uniform comoving
hypersurfaces, ${\cal H} ={a^\prime \over a}$ and the prime
denotes differentiation with respect to the conformal time $\eta$.
In the momentum space, the equation  of motion of $v_k$ is \be
v_k^{\prime\prime}+\left(k^2 -{z^{\prime\prime}\over z}\right)v_k
=0  .\ee where \be {z^{\prime\prime}\over z} ={1\over \eta^2
}\left(\nu^2 -{1\over 4}\right) \ee and $ \nu \simeq {3\over 2}+
2\epsilon_{\rm pha} -\delta_{\rm pha} $ to lowest order of
$\epsilon_{\rm pha}$ and $\delta_{\rm pha}$. Thus following the
same steps as normal scalar field inflation, we obtain \be {\cal
P}_s = {1\over |\epsilon_{\rm pha} | m_p^2 }({h\over
2\pi})^2_{k=ah} \ee and its spectrum index is $ n_s \simeq 1-
4\epsilon_{\rm pha} +2\delta_{\rm pha} $ Similarly for tensor
metric perturbation, we obtain \be {\cal P}_t = {1\over
m_p^2}({h\over 2\pi})^2_{k=ah} \ee and its spectrum index is $ n_t
\simeq -2\epsilon_{\rm pha} $

The tensor/scalar ratio can be expressed as a ratio their
contributions to the CMB quadrupole {\it i.e.} $r\equiv
{C_2^{t}\over C_2^{s}}$. The relation between $r$ and ratio of
amplitudes in the primordial power spectrum ${{\cal P}_t \over
{\cal P}_s}$ depends on the background cosmology. For the
currently favored values of $\Omega_m\simeq 0.3$ and
$\Omega_\Lambda \simeq 0.7$, this relation is \cite{KKMR} \be
r\simeq 10 |\epsilon_{\rm pha} | \label{r}\ee to lowest order. To
obtain more insights for the perturbations spectrum of phantom
inflation , it may be useful to divide them into three interesting
classes compared with normal scalar field inflation\footnote{Here
to make compare more convenient, following \cite{DKK}, we use the
classifying standard of normal scalar field inflation. }, {\it
i.e.} large field (chaotic inflation \cite{L1}), small field (new
inflation \cite{LAS}, natural inflation \cite{FFO}) and hybrid
inflation model \cite{L}, see Fig. 2. For the case that the maxima
of phantom potential are not in $\varphi =0$, like large field and
hybrid model, we can carry out a transform $\varphi\rightarrow
\varphi-\varphi_{\rm max}$ for $V(\varphi,\sigma)$, which make the
hybrid exit mechanism mentioned above still valid.

\begin{figure}[t]
\begin{center}
\includegraphics[width=8cm]{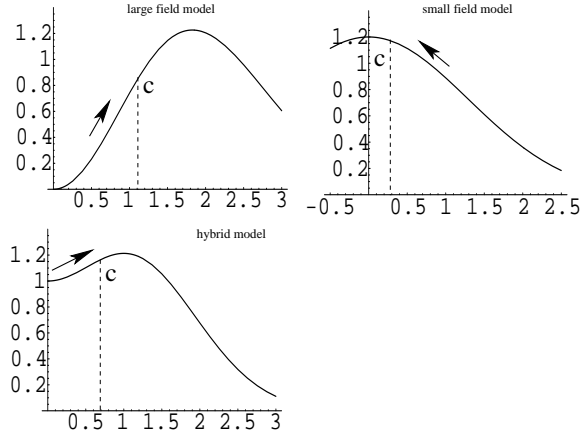}
\caption{The sketch of large field, small field, hybrid model of
phantom inflation, which may be regarded as the cross-section of
the potential of Fig. 1 in $\sigma =0$. For various models, the
phantom field $\varphi$ will climb up along its potential {\it
i.e.} arrowhead direction, when arriving $\varphi_c$, the
additional normal scalar field $\sigma$ becomes unstable and will
roll down along its potential, thus phantom inflation ends. }
\label{fig2}
\end{center}
\end{figure}

{\it Large field models -} characterized by $\epsilon_{\rm pha} <
\delta_{\rm pha} < -\epsilon_{\rm pha}$, in which the phantom
field is driven up to larger value. The generic large field
potentials are polynomial-like potentials $\Lambda ({\varphi\over
\varphi_*})^n$. From (\ref{ed}), we see that when $\varphi \geq
m_p$, the slow-climb conditions are satisfied. Combining
(\ref{ed}) and (\ref{r}), we have \be r={5n\over n+2}(n_s -1) \ee
which is a blue spectrum and reverse with normal large field
inflation models, and when $n\rightarrow \infty$, \be r= 5(n_s-1)
\ee which corresponds exponential potentials $\Lambda
e^{\varphi\over \varphi_*}$. This has been studied in Ref.
\cite{PZ}.

{\it Small field models -} characterized by $\delta_{\rm pha} >
-\epsilon_{\rm pha}$, in which the phantom field is driven up to
smaller value. The generic small field potentials are $\Lambda
(1-({\varphi\over \varphi_*})^n)$, which may be regarded as a
lowest order Taylor expansion of an arbitrary potential about the
origin. From (\ref{ed}), we see that when $\varphi \ll \varphi_*$,
the slow-climb conditions are satisfied. Since $ \epsilon_{\rm
pha} \sim ({\varphi\over \varphi_*})^n \delta_{\rm pha}$, thus $|
\epsilon_{\rm pha}|\ll \delta_{\rm pha} $, the spectrum index is
approximately \be n_s -1\simeq 2\delta_{\rm pha}\ee and tensor
perturbation amplitude is suppressed strongly \be r\simeq 5(n_s
-1) {n\over 2(n-1)} ({\varphi\over \varphi_*})^n\ee



{\it Hybrid models
-} characterized by $ \delta_{\rm pha} < \epsilon_{\rm pha} <0 $,
in which the phantom field is driven from the bottom with a
positive vacuum energy. The generic potentials are $\Lambda
(1+({\varphi\over \varphi_*})^n)$. For $\varphi \gg \varphi_*$
the models are the same as large field models, and for $\varphi\ll
\varphi_* $ we have \be \epsilon_{\rm pha}\simeq -{n^2\over 16\pi
G\varphi_*^2}({\varphi\over \varphi_*})^{2n-2} \ee \be \delta_{\rm
pha} \simeq {2(n-1)\over n}({\varphi_*\over \varphi})^n \,\,
\epsilon_{\rm pha}\ee Thus the red spectrum can be obtained only
when $|\delta_{\rm pha} |> 2|\epsilon_{\rm pha} |$, {\it i.e.}
${2(n-1)\over n }({\varphi_*\over \varphi})^n >2$, and the blue
spectrum only when ${2(n-1)\over n }({\varphi_*\over \varphi})^n
<2$.

In addition, there are {\it Linear models}, $V(\varphi)\sim
\varphi$, which is the boundary between large field and small
field models and characterized by $\delta_{\rm pha} =
-\epsilon_{\rm pha}$, in which the relation \be r= {5\over 3} (n_s
-1)\ee is satisfied.

\begin{figure}[t]
\begin{center}
\includegraphics[width=8cm]{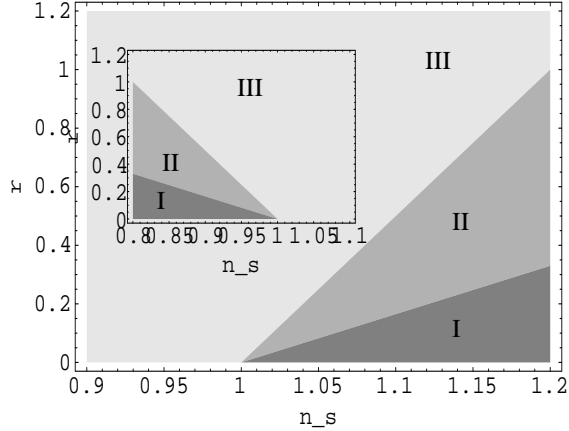}
\caption{The $r  - n_s$ figure for phantom inflation models. {\bf
Inset:} The $r - n_s$ figure for normal scalar field inflation
models \cite{KKMR}. The different shadow Region I II III
correspond the different observational parameter space occupied by
the small field, large field and hybrid models respectively. }
\label{fig2}
\end{center}
\end{figure}

The $r - n_s$ plane figure is a powerful tool showing different
predictions from various inflation models \cite{DKK, KKMR}, which
is plotted in Fig. 3 for phantom inflation, in which various
interesting classes categorized by the slow-climb parameters
completely cover the entire $r - n_s$ plane. For phantom inflation
models, we see that both large field and small field models locate
in blue spectrum ($n_s
>1$) regions, while hybrid models are mainly collected in red
spectrum ($n_s <1$) regions, which is just reverse with that of
normal scalar field inflation models, which is also saw from inset
of Fig. 3 \cite{KKMR}. In fact the definition of slow-climb
parameters $\epsilon_{\rm pha}$ and $\delta_{\rm pha}$ in phantom
inflation is the same as those ($\epsilon_{\rm sca}$ and
$\delta_{\rm sca}$) of normal scalar field inflation, but when
rewriting these parameters in term of potential of field,  \be
\epsilon_{\rm pha} \simeq -{1\over 16\pi G
}({V^{\prime}(\varphi)\over V(\varphi)})^2 \ee \be \delta_{\rm
pha} \simeq {1\over 8\pi G} \left({1\over 2}({V^{\prime}(\varphi)
\over V(\varphi)})^2 -{V^{\prime\prime}(\varphi)\over
V(\varphi)}\right) \ee
the interesting results $\epsilon_{\rm pha}= -\epsilon_{\rm sca}$
and $\delta_{\rm pha} = -\delta_{\rm sca}$ are found, which is the
main reason that the reverse results of $r - n_s$ figure are lead
to and reflects the generic features of phantom inflation.

In summary, we study the various models, exit mechanism and
primordial perturbation spectrum of phantom inflation and discuss
their different features from those of normal scalar field
inflation. We proposed an exit mechanism, in which the phantom
field climbs up along its potential from the bottom of valley, and
when arriving some value, the energy of its potential decays into
radiation by the instability of an additional normal scalar field.
For the perturbation spectrum, phantom inflation models completely
cover the entire $r - n_s$ plane, in which the interesting
constraints are being placed by the observations \cite{Bennett,
Tegmark} and may be tightened in coming years, but with reverse
results from those of normal scalar field inflation.
In some sense of inflation, phantom inflation may be regarded as a
new class different from normal scalar field inflation and worth
studying further.

\textbf{Acknowledgments} We thank Bo Feng, Zong-Kuan Guo, Mingzhe
Li, Xinmin Zhang for helpful discussions. This work is supported
in part by K.C.Wang Postdoc Foundation and also in part by the
National Basic Research Program of China under Grant No.
2003CB716300.

\end{document}